# High precision atom interferometer-based dynamic gravimeter measurement by eliminating the cross-coupling effect

Yang Zhou[1,3], Wenzhang Wang[1,3], Guiguo Ge[1,3], Jinting Li[1,3], Danfang Zhang[1,3], Meng He[1,3], Biao Tang[1,4,5], Jiaqi Zhong[1,4,5], Lin Zhou[1,4,5], Runbing Li[1,4,5], Lin Mao[2], Hao Che[2], Leiyuan Qian[2], Yang Li[2], Fangjun Qin[2,*], Jie Fang[1,4,5]*, Xi Chen[1,4,5]*, Jin Wang[1,4,5], Mingsheng Zhan[1,4,5]

[1] Innovation Academy for Precision Measurement Science and Technology, Chinese Academy of Sciences, Wuhan 430071, China
[2] School of Electrical Engineering, Naval University of Engineering, Wuhan 430033, China
[3] University of Chinese Academy of Sciences, Beijing 100049, China
[4] Wuhan Institute of Quantum Technology, Wuhan 430206, China
[5] Hefei National Laboratory, Hefei 230094, China
*Correspondence: chenxi@apm.ac.cn (X.C.); fangjie@apm.ac.cn (J.F.), haig2005@126.com (F. J. Q.)

**Abstract:** A dynamic gravimeter with an atomic interferometer (AI) can perform absolute gravity measurements with high precision. AI-based dynamic gravity measurement is a type of joint measurement that uses AI sensors and a classical accelerometer. The coupling of the two sensors may degrade the measurement precision. In this study, we analyzed the cross-coupling effect and introduced a recovery vector to suppress this effect. We improved the phase noise of the interference fringe by a factor of 1.9 by performing marine gravity measurements using an AI-based gravimeter and optimizing the recovery vector. Marine gravity measurements were performed, and high gravity measurement precision was achieved. The external and inner coincidence accuracies of the gravity measurement are ±0.42 mGal and ±0.46 mGal, which were improved by factors of 4.18 and 4.21 by optimizing the cross-coupling effect.

**Keywords**：Atom interferometer, Dynamic gravimeter measurement, High precision, Gravimeter, Marine gravity survey, Cold atom

## 1.Introduction

Gravity measurements have important applications in various fields, such as geodesy, geophysics[1], navigation[2], and fundamental physics tests [3,4,5]. Gravity can be obtained from static, dynamic, and satellite measurements. Dynamic gravimeters rely on dynamic carriers, such as ships and aircraft. They obtain accurate and efficient gravitational information and are used as the relative sensors. However, they suffer from drift and must occasionally be calibrated. Recently, dynamic gravimetry based on atomic interferometry has been developed[6,7]. They can measure the value of absolute gravity without measurement drift, which has wide potential applications.

The time-pulsed atom interferometer (AI) was first realized in 1991[8] and has been widely used for precision gravity measurement [9-17], gravity gradient measurements [18,19], and rotation measurements [20,21]. Dynamic gravity measurements based on AI have been realized in moving elevators [22], vehicle [23-26], aircraft [27,28] and ship [7,29,30,31], and the best measurement precision is better than 1 mGal [7,30]. Various methods, such as vibration compensation [32-34] and data filtering [35-38], have been proposed to further improve the precision of AI-based dynamic gravimeters.

An AI-based dynamic gravimeter is a type of joint measurement using AI sensors and a classical accelerometer. The output of a classical accelerometer was compared and corrected using the gravity measurement of the AI in real time. The bias and drift of a classical accelerometer were eliminated to provide accurate and continuous gravity outputs. However, a dynamic environment can degrade the precision of gravity measurement. First, it can affect the trajectory of the cold atom cloud during interference. This changes the Rabi frequencies of the Raman pulses and the fluorescence detection coefficient, decreasing the precision of the gravity measurement. Second, additional noise is induced if the direction of acceleration measured by the classical accelerometer is different from the direction of acceleration felt by the AI. This is called the cross-coupling effect. This effect could be induced by the installation error between the classical accelerometer and the AI, and the crosstalk of different sensing axes of the classical accelerometer [33].

In this study, we analyzed the cross-coupling effect and developed a method to eliminate it. Subsequently, we performed marine gravity measurements and achieved a high measurement precision. The remainder of this paper is organized as follows: In Section 2, we introduce the cross-coupling effect and analyze the induced phase noise. A recovery vector was proposed to eliminate this phase noise. Section 3 introduces an experiment on marine gravity measurements using an AI-based gravimeter. In Section 4, conclusions and discussion are presented.

## 2 Theoretical methods
### 2.1 Joint gravity measurement process

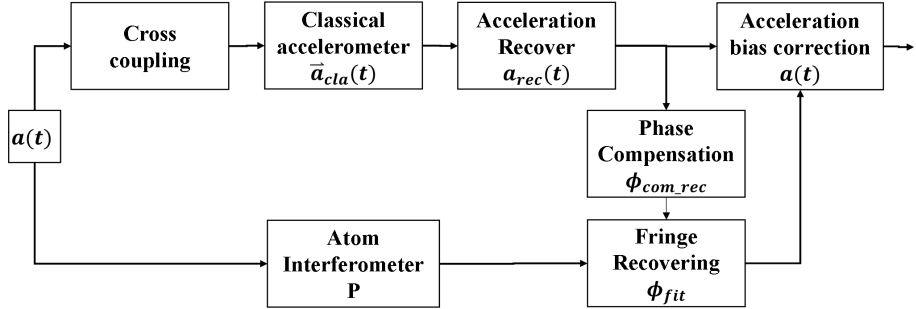

Figure 1. Principle of the joined gravity measurement and the introduction of the cross-coupling effect.

The principle of the joint gravity measurement process is illustrated in Fig. 1. During each gravity measurement, the atomic population $P$ of the AI and the acceleration $a_{cla}(t)$ of the classical accelerometer were obtained. Then, the compensation phase $\phi_{com}$ can be calculated using $a_{cla}(t)$ and the sensitivity function of AI. After several gravity measurements, a series data of $P$ and $\phi_{com}$ can be obtained. The fringe of AI can be recovered by setting $\phi_{com}$ and $P$ as the x- and y-axis coordinates, and the fitting phase $\phi_{fit}$ can be obtained by fitting the fringe with a cosine function. We define the fitting acceleration as $a_{fit} \equiv \phi_{fit}/k_{eff}T^2$, where $k_{eff}$ is the effective wave vector, and T is the interference time. This acceleration represents the bias of classical accelerometers. The absolute acceleration a(t) is calculated as

$$a(t) = a_{cla}(t) - a_{fit} + a_{chirp},  \qquad (1)$$

where $a_{chirp} = 2\pi\alpha_0/k_{eff}$ is the equivalent acceleration induced by the chirp rate $\alpha_0$ of AI. In addition to the gravity acceleration $a_{gra}(t)$, a(t) also contains the motion acceleration $a_{mot}(t)$ of the gravimeter, which can be calculated using the time-varying latitude, longitude, and altitude of the gravimeter[28]. Gravitational acceleration $a_{gra}(t)$ can be calculated as

$$a_{gra}(t) = a_{cla}(t) - a_{fit}(t) + a_{chirp} - a_{mot}(t). \tag{2}$$

The precision of $a_{cla}(t)$ was determined using a classical accelerometer. $a_{chirp}$ is the constant acceleration. The precision of $a_{mot}(t)$ is determined by the precision of the position measurements. $a_{fit}(t)$ is the most relevant term for the precision of joint gravity measurement.

## 2.2 Noise induced by the cross-coupling effect

Assuming that the AI-felt acceleration is $\vec{a}(t) = \{a_x(t), a_y(t), a_z(t)\}$, and the sensitive axis of the AI is in the z-direction, the phase of the AI can be calculated as

$$\phi_{AI} = \int_{-T}^{T} g(t) a_z(t)\, dt, \tag{3}$$

where g(t) is the sensitivity function and $T$ is the interference time of AI. The measured acceleration of the classical accelerometer is $\vec{a}_{cla}(t) = \{a_{cla,x}(t), a_{cla,y}(t), a_{cla,z}(t)\}$. Considering the cross-coupling effect, $\vec{a}_{cla}(t)$ and $\vec{a}(t)$ exhibit the following relationships:

$$\vec{a}_{cla}(t) = \mathbf{C} \cdot \vec{a}(t) + \vec{a}_{off}(t), \tag{4}$$

where $\mathbf{C} = [C_{i,j}]\ (i,j = x, y, z)$ is the coupling matrix and $\vec{a}_{off}(t) = \{a_{off,x}(t), a_{off,y}(t), a_{off,z}(t)\}$ are the measurement offsets of the classical accelerometer. If we use the z-component of the classical accelerometer $a_{cla,z}(t)$ to calculate the compensation phase, we obtain

$$\phi_{com} = \int_{-T}^{T} g(t) a_{cla,z}(t)\, dt. \tag{5}$$

The differential phase between $\phi_{com}$ and $\phi_{AI}$ is equivalent to the fitting phase of the recovered interference fringe.

$$\phi_{fit} = \int_{-T}^{T} g(t) [\vec{c} \cdot \vec{a}(t) + a_{off,z}(t) - a_z(t)]\, dt, \tag{6}$$

where $\vec{c} \equiv \{C_{z,x}, C_{z,y}, C_{z,z}\}$ is the coupling vector. If $\vec{c} \neq \{0,0,1\}$, the acceleration noise will lead to the phase noise of the recovered fringe.

## 2.3 Introduce of the recover vector

To reduce the noise of $\phi_{fit}$, we inserted a recover process before the calculation of the compensation phase, as shown in Fig. 1. We introduced the matrix $\mathbf{D} = [D_{i,j}]\ (i,j = x, y, z)$ to recover the acceleration.

$$\vec{a}_{rec}(t) = D \cdot \vec{a}_{cla}(t) \tag{7}$$

The compensation phase in Eq. (5) is altered to

$$\phi_{com\_rec} = \int_{-T}^{T} g(t) a_{rec,z}(t)\, dt, \tag{8}$$

If $\mathbf{D} = \mathbf{C}^{-1}$, the fitting phase $\phi_{fit}$ is

$$\phi_{fit} = \int_{-T}^{T} g(t) [a_{rec,z}(t) - a_z(t)]\, dt,$$

$$= \int_{-T}^{T} g(t) [\vec{d} \cdot \vec{a}_{off}(t)]\, dt, \tag{9}$$

where the recover vector is denoted as $\vec{d} \equiv \{D_{z,x}, D_{z,y}, D_{z,z}\}$. In this case, acceleration was not coupled to the fitting phase. For a general form of the recovery matrix $\mathbf{D}$, the fitting phase has the following form:

$$\phi_{fit} = \int_{-T}^{T} g(t)[\vec{d}\cdot\vec{a}_{cla}(t) - a_z(t)]dt. \qquad (10)$$

If the recover vector is set to its optimized value, the fitting phase $\phi_{fit}$ will have a minimum noise. The detailed optimization process is described in Section 3.4.

## 3. Marine gravity measurement experiment
## 3.1 Experiment apparatus

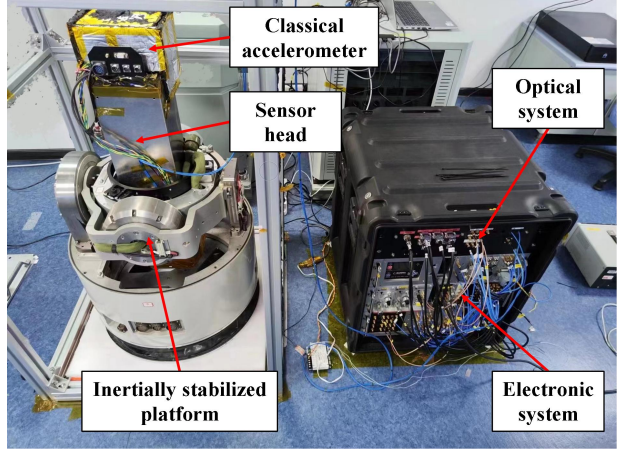

Figure 2. (color online) AI-based dynamic gravimeter for the marine gravity measurement.

We developed a compact AI-based dynamic gravimeter, as shown in Fig. 2. It consists of a sensor head, inertial stabilization platform, optical system, and electronic system [21]. Atomic interference occurs at the sensor head. It utilizes a rubidium-85 cold atom cloud as the test mass, and it is surrounded by a magnetic field shield. Additionally, a classical accelerometer (Titan accelerometer from Nanometrics) is mounted on top of it. The sensor head has a compact size of 17 × 17 × 60 cm and a weight of only 15 kg. It is installed on a dual-axis inertial stabilization platform. The platform offers an angle control accuracy of approximately 0.2 mrad. An optical system is used to provide the required laser power. It consists of several homemade fiber modules and occupies a standard 3U chassis. Two laser beams were sent to the sensor head by two single-mode polarization-maintaining fibers [39,17]. The electronic system was used to drive the components of the gravimeter, generate a time sequence, and acquire and process the experimental data. A GNSS receiver was used to obtain position information.

## 3.2 Systematical error evaluation of the AI-based gravimeter

Before the marine gravity measurements, we performed a systematic error evaluation of the AI-based gravimeter. Long-term gravity measurements were conducted at the National Geodetic Observatory in Wuhan. The interference time was set to T = 15 ms, which was the same as that in the dynamic case. Several system error terms were evaluated, as listed in Table 1. The gravity gradient term was evaluated using the local gravity gradient and height of the sensor head. The single- and double-photon light shift terms were evaluated using the sideband ratio of the Raman laser and the time sequence of the Raman laser pulses. The multi-sideband feature of the Raman laser induces an additional laser line effect because a fiber electro-optic modulator (FEOM) was used to produce the Raman laser. This effect was evaluated using the sideband ratio of the Raman laser, the position of the reflection mirror of the Raman laser, and the trajectory of the cold atom cloud. The solid tide term was evaluated using theoretical calculations. The wave vector inversion method was adopted to suppress the systematic errors induced by the Zeeman and AC Stark shifts. After long time gravity measurement and system error correction, the measured gravity of the AI-based gravimeter and the gravity value of the reference site still had an offset of approximately 110 μGal. This offset may have been caused by the residual Zeeman shift, wavefront aberration of the Raman laser, or other systematic error terms. We treat this offset as a calibration term, as listed in Table 1. We deduced the solid-tide-induced gravity variation from the measured gravity and calculated the Allan standard deviation, as shown in Fig. 3. The gravity measurement resolution was approximately 1.85 mGal and 0.05 mGal at 1 and 5000 s, respectively.

Table 1. Systematical error evaluation for the AI-based gravimeter

| Systematical error terms | Value (mGal) | Uncertainty (mGal) |
|---|---|---|
| Gravity gradient | -0.222 | 0.002 |
| Single photon light shift | 0.000 | 0.008 |
| Double photon light shift | 0.047 | 0.005 |
| Additional laser lines | -0.699 | 0.137 |
| Gravity calibration | -0.116 | 0.050 |
| Calibration in total | -0.990 | 0.147 |

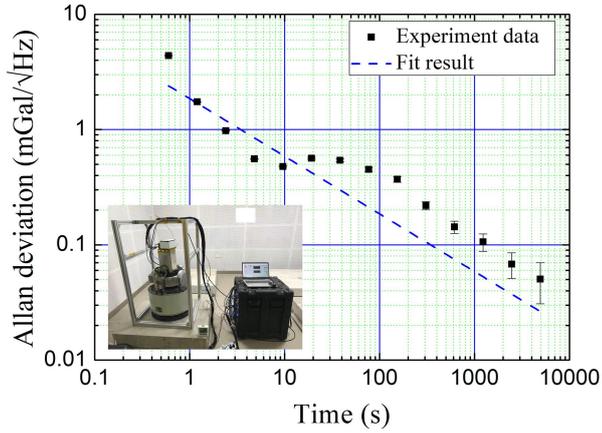

Figure 3. (color online) Allan standard deviation of the measured gravity value at the national geodetic observatory in Wuhan for T = 15 ms.

### 3.3 Gravity comparison under the mooring state

The gravimeter was then transferred from Wuhan to Zhejiang Province and installed on a survey ship. Before and after the dynamic gravity survey, we compared the gravity measurement values with the gravity value of a shore-based gravity reference site in the mooring state. The latitude and height differences between the gravimeter and reference site were measured, and the induced gravity difference was calculated and compensated for the measured gravity. During each comparison, we measured gravity for 40 min and compared the average gravity value with that of the reference site. The measured differences are shown in Fig. 4. All data have a mean value and standard deviation of -0.32 mGal and 0.22 mGal, respectively. No apparent drift was observed before or after dynamic gravity surveying.

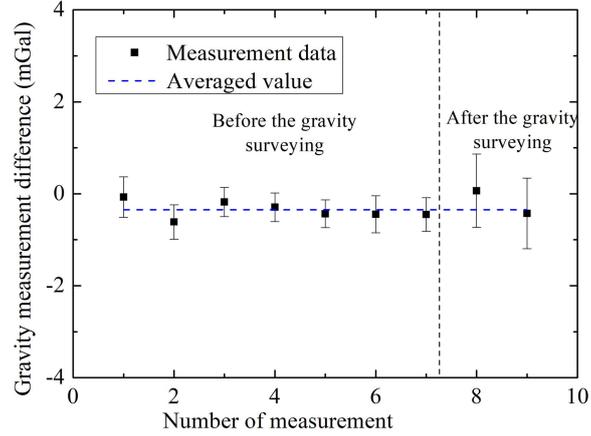

Figure 4. (color online) Gravity comparison with a shore-based gravity reference site under mooring state

### 3.4 Gravity measurement under the sailing state

Marine gravity measurements were conducted in the East China Sea. We carried out repeat survey measurements in the east-west direction. The effect length of a single survey line was 45 km, and the number of survey lines was three. The trajectory of the survey line is shown in Fig. 5(a). The ship's speed was approximately 11 knots during the survey period. A classical shipborne strapdown gravimeter was installed nearby for comparison of gravity measurements. The power spectral density (PSD) of the acceleration measured during the survey and in the mooring state is shown in Fig. 5(b). The accelerations measured along the survey line had peak-to-peak values of approximately 0.6 m/s² peak-to-peak. The interference time of the AI-based gravimeter was set as T = 15 ms during the gravity measurement.

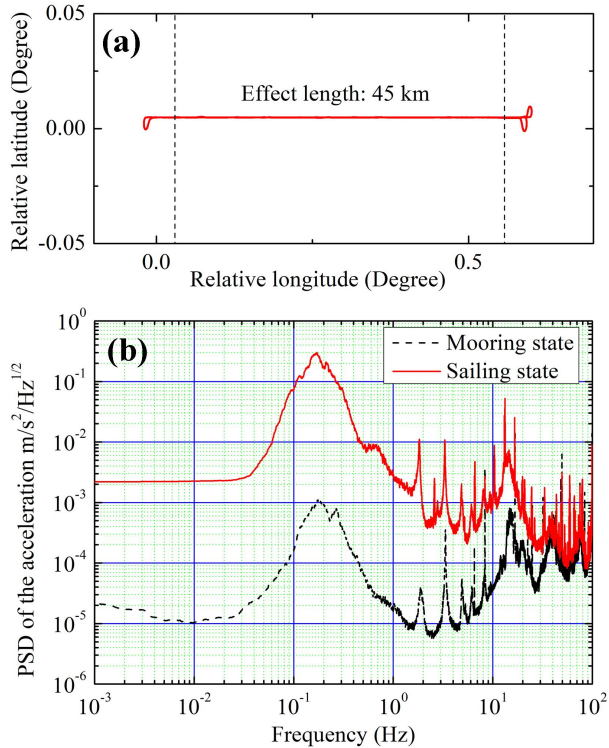

Figure 5. (color online)  (a) The trajectory of the survey line during the marine gravity measurement. (b) The power spectral density amplitude of the measured acceleration in the z direction under the mooring state (black dashed line) and sailing state (red solid line).

Before processing the measured gravity data, we calibrated the recovered vector, as described in Section 2.3. $\phi_{com\_rec}$ and P were set as coordinates of x and y to recover the AI's fringe. The fringe was fitted to obtain the fitting phase $\phi_{fit}$, and the standard deviation $\sigma_{\phi_{fit}}$ of $\phi_{fit}$ for a group of recovered fringes was calculated. We scanned the values of the components of recover vector $\vec{d}$ around {0,0,1} and found the relationship between $\sigma_{\phi_{fit}}$ and recover vector's components. The corresponding curves are referred to as calibration curves, as shown in Fig. 6(a). The curves had a valley shape, and the widths of the valleys were inversely proportional to their corresponding coupling accelerations. This can be understood using Eq. (10). The recovery vector is coupled to the acceleration to introduce phase noise. If the coupled acceleration is small, an offset of the recovered vector from its optimized value will lead to small phase noise. The x-coordinates at the bottom of the valleys represent the optimized values of the components of the recovered vector. The optimized recover vector during the survey measurement is {0.0060, -0.0034, 0.9860}. The uncertainties in the fitting phases of the recovered fringes before and after the optimization process were compared. The phase uncertainties of the fringe for the optimized recovery vector and d = {0, 0, 1} are 0.10 and 0.19, respectively, as shown in Figs. 6(c) and 6(b). The phase noise of the interference fringe improved by a factor of 1.9.

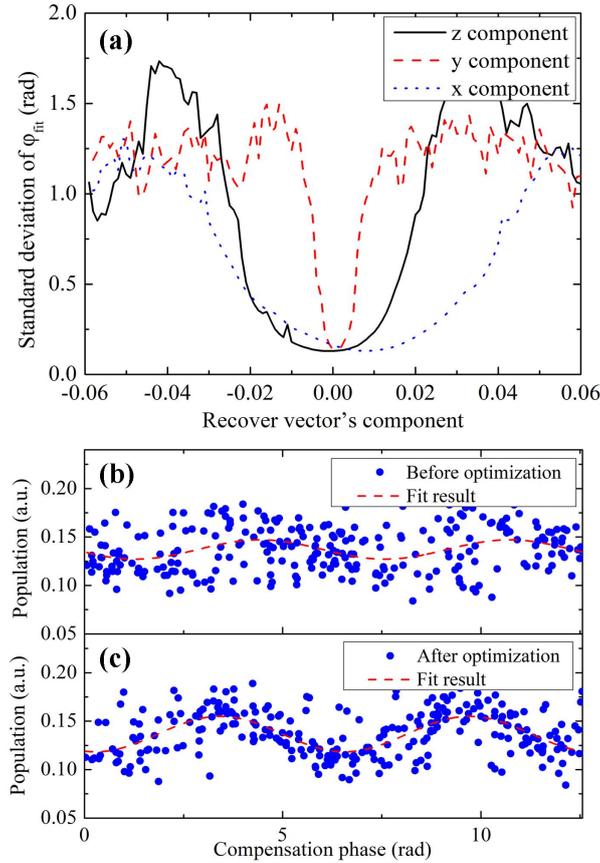

Figure 6. (color online)  (a) The calibration curves for the recover vector during the gravity survey measurement, the z component of recover vector is subtracted by 1 for the convenience of display.

(b) The recovered fringe when the recover vector is set to {0,0,1}. (c) The recovered fringe when the recover vector is set to its optimized value {0.0060, -0.0034, 0.9860}.

After the optimization of the recovery vector, we calculated the gravity anomalies along the survey lines. The data were processed as follows. First, the recover vector was substituted into Eq. (9) to calculate the compensation phase $\phi_{com\_rec}$, and this phase and the atom population P were used to recover the interference fringe. The fitting phase $\phi_{fit}$ and its corresponding acceleration $a_{fit}$ were obtained by fitting the phase of the fringe. This acceleration was used to eliminate the offset of the output of the classical accelerometer $a_{cla}(t)$ to obtain the absolute acceleration $a(t)$, as shown in Eq. (1). The value of measured $a(t)$ during the gravity survey is shown in Fig. 7(a). Second, $a(t)$ is filtered using a fourth-order Bessel low-pass filter to filter the motion acceleration of the surveying ship. The time constant of the filter was set to 300 s, and the filtered acceleration is shown in Fig. 7(b). Third, the Eotvos acceleration, $a_{mot}(t)$, was calculated using the recorded GNSS signal. The same low-pass filter for $a(t)$ was used for this acceleration. The filtered accelerations are shown in Fig. 7(c). The solid tide-induced acceleration and normal gravity were also calculated and subtracted from a(t). The calculated gravity anomaly along the survey lines is shown in Fig. 7(d). This gravity anomaly was compared with that measured using a classical shipborne strapdown gravimeter. The external coincidence accuracies of the gravity anomaly measurements [Appendix A] for the three survey lines were calculated. The results for the three lines are ±0.46 mGal, ±0.42 mGal, and ±0.41 mGal, respectively. The result for the three lines in total is ±0.43 mGal. However, If the recover vector is set to be {0,0,1}, the calculated external coincidence accuracy of the three lines was found to be ±1.80 mGal. The gravity measurement accuracy was improved by a factor of 4.18 by optimizing the cross-coupling effect.

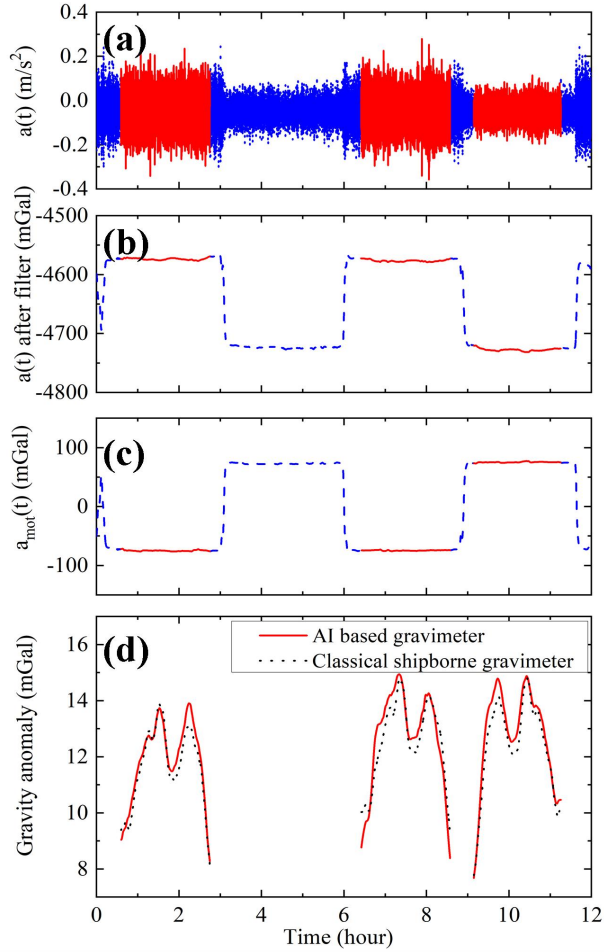

Figure 7. (color online) Data processing process of the gravity anomaly. The red solid line represents data of the three survey lines. (a) The absolute acceleration a(t). (b) The absolute acceleration a(t) after the low pass filter. (c) The calculated Eotvos acceleration $a_{mot}(t)$ after the low pass filter. (d) The measured gravity anomaly of the AI-based gravimeter (red solid line) and the classical shipborne strapdown gravimeter (black dot line).

Then, we calculated the inner coincidence accuracy of the gravity measurements of the three survey lines, as shown in Appendix A. The time-varying gravity anomaly data were converted to position-varying data along the survey lines. The results are shown in Fig. 8(a). Significant gravity measurement deviations were observed across the three survey lines. This was not mainly caused by the measurement offset of the AI-based gravimeter but by the fluctuations in the sea surface height of the three survey lines. To eliminate this effect, the water depth was measured for the three survey lines in real time, and the height-induced gravity variation was calculated, as shown in the insert figure of Fig. 8(b). This gravity variation was deduced from measured gravity anomalies. Therefore, the measured gravity anomaly is transferred from the surface to the bottom of the sea. Then, the gravity anomalies of the three survey lines were compared, and the result are shown in Fig. 8(b). The consistency of the gravity anomalies is better than that in Fig. 8(a), and the inner coincidence accuracy is calculated as ±0.46 mGal. Again, If the recover vector is set to be {0,0,1}, the calculated inner coincidence accuracy was found to be 1.94 mGal. The gravity measurement accuracy was improved by a factor of 4.21.

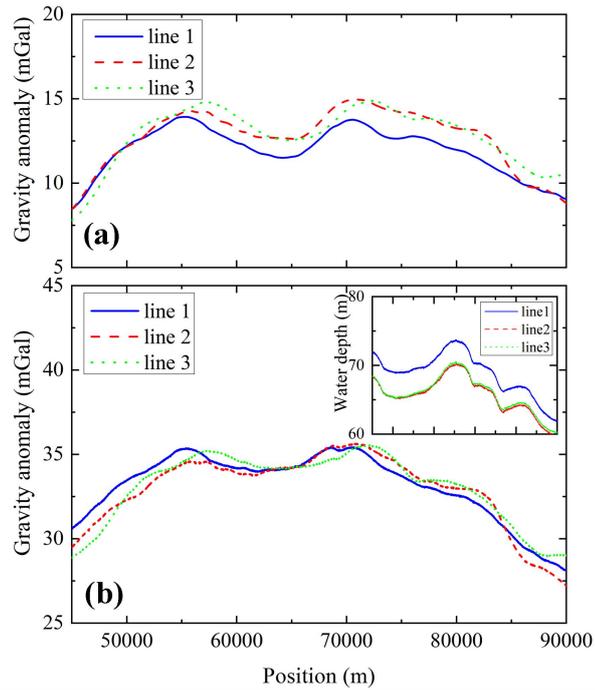

Figure 8. (color online) Comparing of the gravity anomaly measurement during the three survey lines. (a) Before deducting the sea surface height induced gravity. (b) After deducting the sea surface height induced gravity. The insert figure is the measured water depth during the three survey lines.

## 4. Conclusion

In this study, we introduced a model of a joint gravity measurement process for AI-based dynamic gravity measurements. The cross-coupling effect was analyzed, and a recovery vector was introduced to suppress this effect. The phase noise of the interference fringe was improved by a factor of 1.9 in the sailing state by optimizing the value of the recovery vector. Subsequently, the design of an AI-based dynamic gravimeter was introduced, which was used for marine gravity measurement. Before the gravity survey, the systematic error of the AI-based gravimeter was evaluated at a gravity observatory. A gravity comparison with the shore-based gravity reference was performed in the mooring state. The measured gravity difference had a mean value of -0.32 mGal and a standard derivation of 0.22 mGal. Marine gravity measurements were performed using repeated survey lines. The effective length of a single survey line was 45 km, and the number of survey lines was three. The measured gravity anomaly was compared with that of a classical shipborne strapdown gravimeter. After the recover vector optimizing, we achieved high precision for the dynamic gravity measurement. The external coincidence accuracies of the three survey lines in total was ±0.43 mGal. The gravity measurement inner coincidence accuracy of the three survey lines was ±0.46 mGal after considering the water depth induced gravity variation. By optimizing the cross-coupling effect, the gravity measurement external and inner coincidence accuracies were improved by factors of 4.18 and 4.21.

The introduction and optimization of the recovery vector are important for high-precision marine gravity measurements. We believe that the strategies presented in this study will be beneficial for the future design and data analysis of AI-based dynamic gravimeters. Further improvements to the dynamic gravity measurement precision include the accurate calibration of the transfer function of the classical accelerometer and the analysis of the cold atom cloud trajectory under dynamic environments.

**Data availability**

Data supporting the findings of this study are available from the corresponding author upon request.

**Acknowledgements**

We thank the Ocean Survey Department for providing the surveying ship and for assistance in calibrating the gravity values of the gravimeter. This work was supported by the Second Batch of Scientific Experiment Project of the Space Engineering Application System of the China Space Station, Technological Innovation 2030, "Quantum Communication and Quantum Computer" Major Project (2021ZD0300603, 2021ZD0300604), Hubei Provincial Science and Technology Major Project (ZDZX2022000001), China Postdoctoral Science Foundation (2020M672453), and National Natural Science Foundation of China (91536221, 12204493, 42274013).

**AUTHOR CONTRIBUTIONS**

X.C. and Y. Z. built the model and proposed the optimization methods in this study. J.F., X.C., Y.Z., and W.Z.W. implemented the AI-based dynamic gravity measurements. F.J.Q., L.M. H. C, Y. L., and L.Y.Q. were responsible for the calibration and gravity measurements of the classical shipborne gravimeter. X.C. and Y.Z. processed the dynamic gravity data. X.C., F. J. Q., J.F., D.F.Z., G.G.G., J.T.L., M.H., B.T., R.B.L., and J.Q.Z. designed and developed the Al-based gravimeter. M.S.Z. and J. W. provided support and guidance. X.C., Y.Z. and J.F. prepared the manuscript. All the authors have read and approved the final manuscript.

**Competing Interests**

The authors declare no competing interests.

**Reference**

[1] M. Van Camp, O. de Viron, A. Watlet, B. Meurers, O. Francis, C. Caudron, "Geophysics from terrestrial time-variable gravity measurements," Rev. Geophys., vol. 55, no. 4, pp. 938–992, Dec. 2017, DOI. 10.1002/2017RG000566.

[2] H.B. Wang, L. Wu, H. Chai, H. Hsu, Y. Wang, "Technology of gravity aided inertial navigation system and its trial in South China Sea," IET Radar Sonar Navig., vol.10, no. 5, pp. 862-869, Jun. 2016, DOI. 10.1049/iet-rsn.2014.0419.

[3] L. Zhou, C. He, S. T.Yan, X. Chen, D. F. Gao , W. T. Duan, Y. H.Ji,R. D. Xu, B. Tang, C. Zhou , S. Barthwa, Q. Wang, Z. Hou , Z.Y. Xiong, Y. Z. Zhang, M. Liu, W. T. Ni , J. Wang , M. S. Zhan, "Joint mass-and-energy test of the equivalence principle at the $10^{-10}$ level using atoms with specified mass and internal energy," Phys. Rev. A, vol. 104, no. 2, pp. 022822, Aug. 2021, DOI. 10.1103/PhysRevA.104.022822.

[4] P. Asenbaum, C. Overstreet, M. Kim, J. Curti, M. A. Kasevich, "Atom-Interferometric Test of the Equivalence Principle at the $10^{-12}$ Level," Phys. Rev. Lett., vol. 125, no. 19, pp. 191101, Nov. 2020, DOI. 10.1103/PhysRevLett.125.191101.

[5] M. He, X. Chen, J. Fang , Q. F. Chen , H.Y. Sun, Y. B. Wang, J. Q. Zhong, L. Zhou, C. He, J. T. Li, D. F. Zhang, G. G. Ge, W. Z. Wang, Y. Zhou, X. Li, X. W. Zhang, L. Qin , Z. Y. Chen, R.D. Xu , Y. Wang, Z. Y. Xiong, J. J. Jiang, Z. D, Cai, K. Li, G. Zheng, W. H. Peng, J. Wang, M. S. Zhan, "The space cold atom interferometer for testing the equivalence principle in the China Space Station," NPJ Microgravity, vol. 9, no. 1, pp. 58, Jul. 2023, DOI. 10.1038/s41526-023-00306-y.

[6] J. Lautier, L. Volodimer, T. Hardin, S. Merlet, M. Lours, F. Pereira Dos Santos, A. Landragin, "Hybridizing matter-wave and classical accelerometers," Appl. Phys. Lett., vol. 105, no. 14, pp. 144102, Oct 2014, DOI. 10.1063/1.4897358.

[7] Y. Bidel, N. Zahzam, C. Blanchard, A. Bonnin, M. Cadoret, A. Bresson, D. Rouxel, M. F. Lequentrec-Lalancette, "Absolute marine gravimetry with matter-wave interferometry," Nat. Commun., vol. 9, pp. 627, Feb. 2018, DOI. 10.1038/s41467-018-03040-2.

[8] M. Kasevich, S. Chu, "Atomic interferometry using stimulated Raman transitions," Phys. Rev. Lett., vol. 67, no. 2, pp. 181-184, Jul. 1991, DOI. 10.1103/PhysRevLett.67.181.

[9] A. Peters, K. Y. Chung, S. Chu, "High-precision gravity measurements using atom interferometry," Metrologia, vol. 38, no. 1, pp. 25-61, Jan. 2001, DOI. 10.1088/0026-1394/38/1/4.

[10] C. Freier, M. Hauth, V. Schkolnik, B. Leykauf, M. Schilling, H. Wziontek, H. G. Scherneck, J. Müller, and A. Peters, "Mobile quantum gravity sensor with unprecedented stability," J. Phys. Conf. Ser., vol. 723, pp. 012050, Jan. 2016, DOI. 10.1088/1742-6596/723/1/012050.

[11] S. Merlet, Q. Bodart, N. Malossi, A. Landragin, F. Pereira Dos Santos, O. Gitlein and L. Timmen, "Comparison between two mobile absolute gravimeters: optical versus atomic interferometers", Metrologia, vol. 47, pp. L9–L11, June 2010, DIO. 10.1088/0026-1394/47/4/L01.

[12] Z. K. Hu, B. L. Sun, X. C. Duan, M. K. Zhou, L. L. Chen, S. Zhan, Q. Z. Zhang, and J. Luo, "Demonstration of an ultrahigh-sensitivity atom-interferometry absolute gravimeter", PRA, vol. 88, pp. 043610, Aug 2013, DIO. 10.1103/PhysRevA.88.043610.

[13] B. Wu, Z. Y. Wang, B. Cheng, Q. Y. Wang, A. P. Xu and Q. Lin, "The investigation of a μGal-level cold atom gravimeter for field applications" Metrologia, vol. 51, pp. 452–458, May 2014, DIO. 10.1088/0026-1394/51/5/452452–458.

[14] C. Y. Li, J. B. Long, M. Q. Huang, B. Chen, Y. M. Yang, X. Jiang, C. F. Xiang, Z. L. Ma, D. Q. He, L. K. Chen, S. Chen, "Continuous gravity measurement with a portable atom gravimeter," Phys. Rev. A, vol.108, no. 3, pp. 032811, Sep. 2023, DOI10.1103/PhysRevA.108.032811.

[15] S. K. Wang, Y. Zhao, W. Zhuang, T. C. Li, S. Q. Wu, J. Y. Feng and C. J. Li, "Shift evaluation of the atomic gravimeter NIM-AGRb-1 and its comparison with FG5X", Metrologia, Vol. 55, pp. 360–365, Mar 2018, DIO. 10.1088/1681-7575/aab637.

[16] P. W. Huang, B. Tang, X. Chen, J. Q. Zhong, Z. Y. Xiong, L. Zhou, J. Wang, M. S. Zhan, "Accuracy and stability evaluation of the 85Rb atom gravimeter WAG-H5-1 at the 2017 international comparison of absolute gravimeters," Metrologia, vol. 56, no. 4, pp. 045012, Aug. 2019, DOI10.1088/1681-7575/ab2f01.

[17] G. G. Ge , X. Chen, J. T. Li, D. F. Zhang, M. He, W. Z. Wang, Y. Zhou, J. Q. Zhong, B. Tang J. Fang, J. Wang, M.S. Zhan, "Accuracy improvement of a compact 85Rb atom gravimeter by suppressing laser crosstalk and light shift," Sensor, vol. 23, no. 13, pp. 6115, Jul. 2023, DOI. 10.3390/s23136115.

[18] C. Janvier, V. Ménoret, B. Desruelle, "Compact differential gravimeter at the quantum projection-noise limit", PRA, vol. 105, pp. 022801, Jan 2022, DIO. 10.1103/PhysRevA.105.022801.

[19] W. Lu, J. Q. Zhong, X. W. Zhang, W. Liu, L. Zhu, W. H. Xu, X. Chen, B. Tang, J. Wang, M. S. Zhan, "Compact high-resolution absolute-gravity gradiometer based on atom interferometers," Phys. Rev. Applied, vol. 18, no. 5, pp. 054091, Nov 2022, DOI. 10.1103/PhysRevApplied.18.054091.

[20] I. Dutta, D. Savoie, B. Fang, B. Venon, C. L. Garrido Alzar, R. Geiger, A. Landragin, "Continuous cold-atom inertial sensor with 1 nrad/sec rotation stability," Phys. Rev. Lett., vol. 116, no. 18, pp. 183003, May 2016, DOI. 10.1103/PhysRevLett.116.183003.

[21] Z. W. Yao, H. H. Chen, S. B. Lu, R. B. Li, Z. X. Lu, X. L. Chen, G. H. Yu, M. Jiang, C. Sun, W. T. Ni, J. Wang, M. S. Zhan, "Self-alignment of a large-area dual-atom-interferometer gyroscope using parameter-decoupled phase-seeking calibrations", PRA, vol. 103, pp. 023319, Feb 2021, DOI. 10.1103/PhysRevA.103.023319.

[22] Y. Bidel, O. Carraz, R. Charrière, M. Cadoret, N. Zahzam, A. Bresson, "Compact cold atom gravimeter for field applications," Appl. Phys. Lett., vol. 102, no. 14, Apr. 2013, pp. 144107, DOI. 10.1063/1.4801756.

[23] X. Wu, Z. Pagel, B. S. Malek, T. H. Nguyen, F. Zi, D. S. Scheirer, H. Müller, "Gravity surveys using a mobile atom interferometer," Sci. Adv., vol. 5, no. 9, pp. eaax0800, Sep. 2019, DOI10.1126/sciadv.aax0800.

[24] H. L. Wang, K. N. Wang, Y. P. Xu, Y. T. Tang, B. Wu, B. Cheng, L. Y. Wu, Y. Zhou, K. X. Weng, D. Zhu, P. J. Chen, K. J. Zhang, Q. Lin, "A truck-borne system based on cold atom gravimeter for measuring the absolute gravity in the field," Sensors, vol. 22, no. 16, pp. 6172, Aug. 2022, DOI. 10.3390/s22166172.

[25] J. Y. Zhang, W. J. Xu, S. D. Sun, Y. B. Shu, Q. Luo, Y. Cheng, Z. K. Hu, M. K. Zhou, "A car-based portable atom gravimeter and its application in field gravity survey," AIP Adv., vol. 11, no. 11, pp. 115223, Nov. 2021, DOI. 10.1063/5.0068761.

[26] J. Guo, S. Q. Ma, C. Zhou, J. X. Liu, B. Wang, D. B. Pan, H. C. Mao, "Vibration compensation for a vehicle-mounted atom gravimeter," IEEE Sens. J., vol. 22, no. 13, pp. 12939-12946, Jul. 2022, DOI10.1109/JSEN.2022.3179297.

[27] R. Geiger, V. Ménoret, G. Stern, N. Zahzam, P. Cheinet, B. Battelier, A. Villing, F. Moron, M. Lours, Y. Bidel, A. Bresson, A. Landragin, P. Bouyer, "Detecting inertial effects with airborne matter-wave interferometry," Nat. Commun., vol. 2, pp. 474, Sep. 2011, DOI. 10.1038/ncomms1479.

[28] Y. Bidel, N. Zahzam, A. Bresson, C. Blanchard, M. Cadoret, A. V. Olesen, R. Forsberg, "Absolute airborne gravimetry with a cold atom sensor," J. Geod., vol. 94, no. 2, pp. 20, Jan. 2020, DOI. 10.1007/s00190-020-01350-2.

[29] B. Wu, C. Zhang, K. N. Wang, B. Cheng, D. Zhu, R. Li, X. L. Wang, Q. Lin, Z. K. Qiao, Y. Zhou, "Marine absolute gravity field surveys based on cold atomic gravimeter," IEEE Sens. J., to be published, DOI. 10.1109/JSEN.2023.3309499.

[30] Z. K. Qiao, P. Yuan, J. J. Zhang, Z. Y. Zhang, L. L. Li, D. Zhu, M. R. Jiang, H. Y. Shi, R. Hu, F. Zhou, Q. Y. Wang, Y. Zhou, B. Wu, Q. Lin, "Error analysis and filtering methods for absolute ocean gravity data," IEEE Sens. J., vol. 23, no. 13, pp. 14346-14355, Jul. 2023, DOI. 10.1109/JSEN.2023.3272551.

[31] H. Che, A. Li, J. Fang, G. G. Ge, W. Gao, Y. Zhang, C. Liu, J. N. Xu, L. B. Chang, C. F. Huang, W. B. Gong, D. Y. Li, Xi Chen, F. J. Qin, "Ship-borne dynamic absolute gravity measurement based on cold atom gravimeter," Acta Phys. Sin., vol. 71, no. 11, pp. 113701, Jun. 2022, DOI. 10.7498/aps.71.20220113.

[32] J. Le Gouët, T. Mehlstäubler, J. Kim, S. Merlet, A. Clairon, A. Landragin, and F. Pereira Dos Santos, "Limits to the sensitivity of a low noise compact atomic gravimeter," Appl. Phys. B, vol. 92, no. 2, pp. 133–144, Aug. 2008, DOI. 10.1007/s00340-008-3088-1.

[33] S. Merlet, J. le Gouët, Q. Bodart, A. Clairon, A. Landragin, F. Pereira Dos Santos, P. Rouchon, "Operating an atom interferometer beyond its linear range," Metrologia, vol. 46, no. 1, pp. 87–94, Feb. 2009, DOI. 10.1088/0026-1394/46/1/011.

[34] H. Che, A. Li, Z. Zhou, W. B. Gong, J. X. Ma, F. J. Qin, "An approach of vibration compensation for atomic gravimeter under complex vibration environment," Sensors, vol. 23, no. 7, pp. 3535, DOI. 10.3390/s23073535.

[35] P. Cheiney, L. Fouché, S. Templier, F. Napolitano, B. Battelier, P. Bouyer, B. Barret, "Navigation-compatible hybrid quantum accelerometer using a Kalman filter," Phys. Rev. Appl., vol. 10, no. 3, pp. 034030, Sep. 2018, DOI. 10.1103/PhysRevApplied.10.034030.

[36] D. Zhu, H. Xu, Y. Zhou, B. Wu. B. Cheng, K. N. Wang, P. J. Chen, S. T. Gao, W. K. Weng, H. L. Wang, S. P. Peng, Z. K. Qiao, L. X. Wang, Q. Li, "Data processing of shipborne absolute gravity measurement based on the extended Kalman filter algorithm," Acta Phys. Sin., vol. 71, no. 13, pp. 133702, Jul. 2022, DOI. 10.7498/aps.71.20220071.

[37] H. Che, A. Li, J. Fang, X. Chen, F. J. Qin, "Interference fringe fitting of atom gravimeter based on fitness particle swarm optimization," AIP Adv., vol. 12, no. 7, pp. 075211, Jun 2022, DOI. 10.1063/5.0096967.

[38] C. F. Huang, A. Li, F. J. Qin, J. Fang, X. Chen, "An atomic gravimeter dynamic measurement method based on Kalman filter," Meas. Sci. Technol., vol. 34, no. 1, pp. 015013, Jan. 2023, DOI. 10.1088/1361-6501/ac8e8b.

[39] J. Fang, J. G. Hu, X. Chen, H. R. Zhu, L. Zhou, J. Q. Zhong, J. Wang, M. S. Zhan, "Realization of a compact one-seed laser system for atom interferometer-based gravimeters" OPTICS EXPRESS, vol. 26, no. 2, pp. 1586, Jan 2018, DIO. 10.1364/OE.26.001586.

**APPENDIX A: Define the inner and external coincidence accuracy**

The inner coincidence accuracy is given by

$$M_{inn} = \pm \sqrt{\frac{\sum_{i=1}^{n}\sum_{j=1}^{m}(g_{ij}-g_i)^2}{n\times(m-1)}}, \qquad (A\text{-}1)$$

where m is the number of survey lines, n is the number of data points for each survey line, $g_{ij}$ is the gravity anomaly data of the survey lines, and $g_i = \sum_{j=1}^{m} g_{ij}/m$ is the averaged value of the gravity anomaly along the survey line.

The external coincidence accuracy is expressed as

$$M_{ext} = \pm \sqrt{\frac{\sum_{i=1}^{n}(g_{AI,i}-g_{Cla,i})^2}{n}}, \qquad (A\text{-}2)$$

where n is the number of data points on each survey line, $g_{AI,i}$ is the gravity anomaly measured by the AI-based gravimeter, and $g_{Cla,i}$ is the gravity anomaly measured by the classical shipborne strapdown gravimeter.